\documentclass[aps,prl,preprint]{revtex4-1}
\usepackage{graphicx}
\usepackage{amsfonts}
\usepackage{amsmath}
\usepackage{color}
\usepackage{bm}

\newcommand{\beq}{\begin{equation}}
\newcommand{\eeq}{\end{equation}}

\begin{document}

\title{Collisional decoherence of superposition states in an ultracold gas near a Feshbach resonance}

\author{Jie Cui}
\affiliation{Department of Chemistry, University of British Columbia, Vancouver, B.C., V6T 1Z1, Canada}

\author{Roman V. Krems}
\affiliation{Department of Chemistry, University of British Columbia, Vancouver, B.C., V6T 1Z1, Canada}
\pacs{}
\date{\today}

\begin{abstract}
We present expressions demonstrating that collisional decoherence of ultracold atoms or molecules in a coherent superposition of non-degenerate quantum states
is suppressed when both the real and imaginary parts of the scattering lengths for the states in the coherent superposition are equal. We show that the rate of collisional decoherence can be enhanced or suppressed by varying an external magnetic  field near a Feshbach resonance. For some resonances, the suppression is very dramatic.  We propose a method for measuring the scattering length of ultracold particles in excited quantum states exhibiting Feshbach resonances. 

\end{abstract}

\maketitle

\section{Introduction}

Quantum coherence between internal states of atoms and molecules is exploited for many diverse applications, including Ramsey interferometry \cite{interferometry}, quantum computing \cite{quantum-computing}, quantum simulation of spin-lattice Hamiltonians \cite{zoller} and exciton energy transfer \cite{exciton}, coherent control of intra- and inter-molecular dynamics \cite{coherent-control}, magnetometry \cite{magnetometry}, laser-field alignment of molecules \cite{alignment} and stereochemistry \cite{stereochemistry}. These applications are always limited by decoherence. Decoherence arises from random forces exerted on atoms or molecules in coherent superposition states. Decoherence underlies the quantum-classical system transition \cite{decoherence-book} and understanding decoherence mechanisms is key to unraveling the relationship between quantum and classical mechanics \cite{zurek}. Molecular collisions is  an important source of decoherence in experiments with atoms and molecules in the gas phase. Due to the stochastic nature of molecular collisions, collisional decoherence is usually uncontrollable. This limits the experimental studies of collisional decoherence and other decoherence sources, especially when several decoherence mechanisms have similar timescales.  On the other hand, collisional decoherence can be used as a probe of interparticle interactions in a gas \cite{cross-sections1, cross-sections2}. 

  There is currently great interest in experimental work with atoms and molecules at ultralow temperatures. Cooling molecular gases to subKelvin temperatures eliminates many complications of the experimental measurements at ambient temperatures and opens up the possibility of controlling intermolecular interactions \cite{njp-review, coldmolecules-book}.  This suggests a possibility of designing a system with tunable collisional decoherence. The most widely used technique for controlling interparticle interactions in ultracold atomic gases is based on tuning a Feshbach resonance with an external magnetic field \cite{feshbach}. The collision properties of ultracold atoms in a particular quantum state are determined by two parameters: the real and imaginary parts of the scattering length. Feshbach resonances modify the scattering length, allowing for the magnetic field control of atomic collisions \cite{feshbach}. 
In the present work, we consider an ensemble of ultracold particles (atoms or molecules) prepared in a coherent superposition of two quantum states and explore the possibility of using a Feshbach resonance to suppress collisional decoherence. With two states in the coherent superposition, the decoherence is determined by the real and imaginary parts of two scattering lengths. We present numerical calculations that demonstrate the possibility of tuning these four parameters to the regime where the damping of coherent oscillations must be dramatically suppressed. 

  The second problem we consider here is the possibility of using an experimental measurement of collisional decoherence for determining the scattering length of ultracold atoms or molecules. The scattering length of ultracold atoms can be determined by measuring the thermalization rate of an ultracold gas driven out of thermal equilibrium. This method can only be used for atoms in the absolute ground state immune to inelastic collisions. 
An alternative approach is based on mapping out the positions and widths of Feshbach resonances by measuring three-body recombination and fitting the experimental data by a model based on multichannel scattering calculations \cite{feshbach,zhiying}. This method can provide the scattering lengths of atoms in different internal states, but the extension of this approach to molecular systems is anticipated to be extremely difficult. Clearly, there is a need for the development of a versatile experimental technique for measuring the scattering length of ultracold atoms and molecules. We present an analysis demonstrating that, if the scattering length for any state is known, the magnetic field dependence of the scattering length of any other state can be determined by measuring the depletion of coherence. 

\section{Theory}

The theoretical analysis presented here is general for any atomic or molecular system. 
  For concreteness, we consider an ensemble of ultracold molecules prepared in a coherent superposition of two non-degenerate states, labeled $a$ and $b$,  and placed in an ultracold buffer gas of structureless atoms, such as Mg($^1S$). The buffer gas represents the environment that destroys coherence between molecular states. The coupling to the environment is enabled by the atom - molecule collisions. The microscopic theory of collisional decoherence due to molecular collisions was presented in our earlier work \cite{chris-decoherence}. It is assumed that molecules leave the buffer gas when they undergo inelastic collisions. The inelastic collisions therefore deplete the molecular ensemble. Following Ref. \cite{chris-decoherence}, we describe the decoherence by two measurable quantities,
the decay of the off-diagonal elements of the density matrix (the total coherence signal) 

\begin{eqnarray}
|\rho_{a, b}^{\mathrm{el}}(t)|&=&\sqrt{\rho_{a, b}^{\mathrm{el}}(t)\rho_{a,b}^{\mathrm{el}}(t)}
\label{total-rho}
\end{eqnarray}

\noindent
and the relative coherence describing the coherence of molecules remaining in the gas after inelastic collisions,

\begin{eqnarray}
  \eta_{ab} = \left ( \frac{\rho^{\rm el}_{a, b} \rho^{\rm el}_{b, a}}{\rho^{\rm el}_{a, a}\rho^{\rm el}_{b, b} } \right )^{1/2},
\label{eta}
\end{eqnarray}

\noindent
where the matrix elements $\rho^{\rm el}_{ij}$ are defined as 

\begin{eqnarray}
\rho^{\rm el}_{ij} = \int {\rm tr}_{\rm gas}{\left [ S_{ii} (\hat{\rho}_{ij} \otimes \hat{\rho}^{\rm gas}) S^\dagger_{jj} \right ]} d^3P, 
\label{rho}
\end{eqnarray}

\noindent
$S_{ii}$ denotes the diagonal elements of the scattering matrix describing collisions of molecules with the buffer gas atoms, $\hat{\rho}_{ij}$ denotes the matrix elements  $\langle i | \hat{\rho} | j \rangle$ of the density operator for the molecules, $\hat{\rho}^{\rm gas}$ is the density operator for the buffer gas atoms and the integration is over the translational momenta of the molecules. At first sight, it may appear that the quantities defined by Eqs. (\ref{total-rho}) and (\ref{eta}) are unaffected by inelastic collisions. However, due to the unitarity of the scattering $S$-matrix, the rates of inelastic scattering affect the diagonal elements of the scattering matrix and hence the density matrix elements defined by Eq. (\ref{rho}). 

  The time dependence of $\eta_{a b}$ is given by \cite{chris-decoherence}
  
\begin{eqnarray}
\eta_{a b}(t) = \eta_{a b}(t = 0) \left[ 1+ T^{1/2} t \xi_1 + T \left( \xi_{2,1} t + \frac{t^2}{2}\xi_{2,2} \right) + \cdots \right], 
\end{eqnarray}

\noindent
where $T$ is the temperature of the buffer gas and $\xi$ are the expansion coefficients defined below. 
The decoherence rate is 

\begin{equation}
\frac{d}{dt} \eta_{a b}(t) = \eta_{ab}(0)\left[ T^{1/2}\xi_1 + T(\xi_{2,1} + t\xi_{2,2}) + \cdots \right].
\label{rate}
\end{equation}

\noindent 
At low temperatures and short times, the decoherence rate is thus determined by the coefficient $\xi_1$, while the coefficients $\xi_{2,1}$ and $\xi_{2,2}$ may become important at larger times and higher temperatures.  The $s$-wave scattering amplitude of ultracold molecules in state $i$ ($f_{ii}$) can be expanded in powers of linear momentum ($p=\hbar k$) as follows: $f_{ii}(p) \approx - a_i + b_i p + c_i p^2$.  Ref. \cite{chris-decoherence} presents the expressions for the coefficients $\xi$ in terms of the expansion coefficients $a_i$ and $b_i$ of the scattering amplitude. It is more convenient for the present study to express the coefficients $\xi$ in terms of the real and imaginary parts of the scattering length for molecules in states $a$ and $b$. To do this, we rewrite the $s$-wave scattering amplitudes $f_{ii}$ as 

\begin{equation}
 \ f_{ii}(k)=\frac{1}{g_i(k^2)-ik}
 \end{equation}

\noindent
and use the well-known expansion of the function $g_i(k^2)$, where the first two terms are determined by the scattering length and the effective range \cite{effective-range}. This provides a relation between the coefficients $b_i$ and the scattering length of molecules in state $i$ and allows us to write the coefficients $\xi_1$, $\xi_{2,1}$ and $\xi_{2,2}$ as follows: 

\begin{eqnarray}
\xi_1&=& - \frac{2^{5/2}\pi^{1/2}n_{\rm{gas}}k_B^{1/2}}{{m^*}^{1/2}}\left[ (\alpha_b - \alpha_{a})^2 + (\beta_b - \beta_{a})^2 \right] ,
\label{xi-1}
\end{eqnarray}
\begin{eqnarray}
\xi_{2,1} &=& 12\pi n_{\rm{gas}}k_B r^{3/2} (\beta_b + \beta_{a} )\left[ (\alpha_b - \alpha_{a} )^2 + (\beta_b - \beta_{a})^2 \right] /\hbar,
\end{eqnarray}
\begin{eqnarray}
\xi_{2,2}  &=&\left\{\frac{32\pi^3n^2_{\rm{gas}}k_B}{m^*} + \frac{8 \pi k_B n^2_{\rm{gas}}}{m}\left[ 3(2r+1)^{1/2} + \frac{1+2r+3r^2}{r} \sin^{-1} \left( \frac{r}{r+1} \right) - 4(1+r) \right] \right\} \nonumber \\
& &\times |a_b - a_{a} |^4 - \frac{64 \pi k_B n^2_{\rm{gas}}}{m} \left[ 3(2r+1)^{1/2} + \frac{1+2r+3r^2}{r} \sin^{-1} \left( \frac{r}{r+1} \right) - 4(1+r) \right] \nonumber \\
& & \times\left(\beta_b^2+\beta_{a}^2+\beta_{b}\beta_{a}\right)|a_b - a_{a} |^2 .
\label{xi-22}
\end{eqnarray}

\noindent
Here, $n_{\rm gas}$ is the density of the buffer gas, $k_B$ is the Boltzmann constant, $m$ is the mass of the buffer gas atom, $m^*$ is the reduced mass of the atom - molecule colliding pair, $r$ is the ratio of the buffer gas atom mass and the molecule mass, $\alpha_i$  and $\beta_i$ denote the real and imaginary parts, correspondingly,  of the scattering length $a_i$ describing ultracold collisions of molecules in state $i$ with buffer gas atoms. 

Eqs. (\ref{xi-1}) - (\ref{xi-22}) lead to two observations: (i) The decoherence rate vanishes if both the real and the imaginary parts of the scattering lengths for the two states $a$ and $b$ have the same magnitude and sign. (ii) If one of the states $a$ or $b$ is the absolute ground state, the corresponding scattering length is real and the decoherence rate can never vanish, even if the real parts of the two scattering lengths are equal. This means that the decoherence of coherent superpositions of states $a$ and $b$ can be suppressed to zero {\it only} if {\it both} of the molecular states are excited states.

\section{Results}

The scattering length of ultracold atoms and molecules can be tuned by an external magnetic field in the presence of scattering Feshbach resonances \cite{njp-review}. Near the resonances, both the real and imaginary parts of the scattering length undergo rapid variation as functions of the external field. Atoms and molecules in different internal states usually exhibit scattering resonances at significantly different magnitudes of external fields. If the scattering lengths of the states $a$ and $b$ were both real, it would be generally easy to find an interval of an external field where the scattering length of state $a$ would vary rapidly and that of state $b$ would be independent of the external field. It would therefore be easy to find an external field magnitude, at which the two scattering lengths would be equal and the decoherence rate would vanish. However, it is not clear if the two scattering lengths can be made the same, if they are both complex.

To explore this, we calculate the decoherence rate for NH molecules prepared in a coherent superposition of different rotational states in a buffer gas of ultracold Mg atoms. The experimental work on cooling NH molecules using buffer gas cooling is currently actively pursued \cite{doyle}. It was also shown theoretically that NH molecules can be effectively cooled if placed in an ultracold gas of Mg atoms \cite{alisdair}. The cross sections for Mg - NH collisions at ulracold temperatures exhibit multiple Feshbach resonances \cite{alisdair}. We use the following parameters for our calculations: $T = 10^{-6}$ K, $n_{\rm gas} = 10^{11}$ cm$^{-3}$, $m = 24.3$ g/mol, and $r = 1.62$. The scattering lengths are computed using the Mg - NH interaction potential presented in Ref. \cite{potential}. The scattering calculations are performed as described in Ref. \cite{jcp-2004}. 

  The energy levels of the NH molecule are separated into manifolds that can be labeled by a (nearly conserved) quantum number $N$ of the rotational angular momentum. 
 In the presence of a magnetic field, each $N$ manifold is split into several fine structure and Zeeman sublevels (see Figure 1).
  Within each manifold, the energy levels can be labeled by the quantum number of the total angular momentum ($j$) and its projection on the magnetic field axis ($m_j$). The total angular momentum is a good quantum number at zero magnetic field. The label $j$ is used for states that adiabatically correlate with a particular $j$-level at zero field.  
   We consider NH molecules prepared in a coherent superposition of the state $(j=4, m_j=-4)$ in the $N=3$ manifold (state $a$) and the state $(j=3, m_j = -3)$ in the $N=2$ manifold (state $b$). We assume that the molecules are initially in a pure state so that $\eta_{a,b}(t=0) = 1$. 
 Figure 2 presents the scattering lengths for ultracold collisions of Mg atoms with NH molecules in these states. Figure 3 shows that both the real and imaginary parts of the scattering lengths become similar at certain magnitudes of the magnetic field and demonstrates that the decoherence rate is dramatically suppressed at certain magnitudes of the magnetic field. The calculation shows that the decoherence rate can be modified over the range spanning six orders of magnitude by varying the magnetic field. In order to verify that the suppression of collisional decoherence observed in Figure 3 is not coincidental, we repeated the calculations for several different pairs of molecular states, labeled in Figure 1. For each combination of states, there is an interval of the magnetic field, where the decoherence rate is suppressed by more than three orders of magnitude.

Eqs. (\ref{xi-1}) - (\ref{xi-22})  show that the decoherence rate is determined by the difference of the scattering lengths in states $a$ and $b$. This suggests that a measurement of the decoherence rates can be used to determine the scattering length of state $b$ if the scattering length of state $a$ is known. To do this, it is necessary to measure the time evolution of both 
$|\rho_{a, b}^{\mathrm{el}}(t)|$ and $\eta_{a,b}(t)$. At low temperatures, the time evolution of $|\rho_{a, b}^{\mathrm{el}}(t)|$ is given by \cite{chris-decoherence}
\begin{eqnarray}
|\rho_{a,b}^{\mathrm{el}}(t)|&=&\sqrt{\rho_{ a,b}^{\mathrm{el}}(t)\rho_{b,a}^{\mathrm{el}}(t)} 
= |\rho_{a,b}(0)|e^{\zeta_0 t}\left[1+T^{1/2}\zeta_1 t +\cdots \right],
\end{eqnarray}
where 
\begin{eqnarray}
\zeta_0& = &-\frac{4\pi\hbar n_{\mathrm{gas}}}{m^*}\frac{(\beta_{a}+\beta_{b})}{2} , \\
\zeta_1&=&\frac{2^{5/2}\pi^{1/2}k_{\mathrm{B}}^{1/2} n_{\mathrm{gas}}}{{m^*}^{1/2}}[(\beta_{a}+\beta_{b})^2-(\alpha_{a}-\alpha_{b})^2] .
\end{eqnarray}
In the limit of low $T$, the decay of $|\rho_{a,b}^{\mathrm{el}}(t)|$ is entirely determined by Eq. (11), which provides the relation between $\beta_a$ and $\beta_b$. 
The magnitudes of $\beta_a$ and $\beta_b$ can also be determined directly by measuring the decay of the populations of states $a$ and $b$.  To obtain $\alpha_b$ from $\alpha_a$, one can use the decoherence rate (\ref{rate})
\begin{eqnarray}
\alpha_{b}=\alpha_{a}\pm\sqrt{\frac{R_{ab}}{C T^{1/2}}-(\beta_{b}-\beta_{a})^2}
\end{eqnarray}
where $R_{ab} = {d}\eta_{a b}(t \rightarrow0)/dt$ and $C =2^{5/2}\pi^{1/2}n_{\rm gas} k_{ B}^{1/2}/{m{^\ast}}^{1/2}$. Unfortunately, this equation is not unambiguous and provides two possible values for $\alpha_b$. This problem arises because the decoherence rate is determined by the absolute difference of the scattering lengths. 

  To overcome this problem, we propose first to create a coherent superposition of state $a$ and another state $d$, whose scattering length has no resonances and is independent of the magnetic field $B$ in a given interval of magnetic fields $\Delta B$, where state $a$ exhibits, at least, one Feshbach resonance. 
Most atomic and molecular systems possess such states. In ultracold alkali metal atoms, hyperfine  states with the largest value of the total angular momentum $f$ and the angular momentum projection $m_f=f$ are resonance-free. In the molecular system considered here, the maximally stretched state labeled $d$ in Figure 1 exhibits no resonances (see Figure 2). The real part of the scattering length $\alpha_d$ is also, in principle, undetermined $\alpha_d = \alpha_a \pm \sqrt{R_{ad}/CT^{1/2} - (\beta_a - \beta_d)^2}$. However, because $\alpha_a$ is an analytic function of the magnetic field \cite{resonance} and $R_{ad}$ is a continuous function of the magnetic field, only one of the signs gives the value of $\alpha_d$ that is independent of the magnetic field. This is illustrated in Figure 4. Thus, given $\alpha_a(B)$ and the decoherence rate $R_{ad}(B)$, it is possible to determine $\alpha_d$ non-ambiguously. Once $\alpha_d$ and $\beta_d$ are known, the scattering length of any other state $x$ exhibiting a Feshbach resonance can be determined non-ambigously by measuring the decoherence rate $R_{ax}$ as a function of the magnetic field. 

In conclusion of this section, we need to point out the challenges of measuring the decoherence rate at ultralow temperatures. We note that  inelastic collisions lead to the decay of all the density matrix elements $\rho_{a, a}^{\mathrm{el}}$, $\rho_{b, b}^{\mathrm{el}}$ and $|\rho_{a, b}^{\mathrm{el}}|$ due to trap loss even if the scattering lengths of the states $a$ and $b$ are equal and there is no decoherence of trapped molecules.  As demonstrated in Figure 3, when the decoherence rate $d\eta_{ab}/dt$ is suppressed, the matrix elements $\rho_{a, a}^{\mathrm{el}}$, $\rho_{b, b}^{\mathrm{el}}$ and $|\rho_{a, b}^{\mathrm{el}}|$ decay at the same rate. This decay limits the lifetime of the experiments.  The decay rates of the diagonal matrix elements are determined by the imaginary parts of the scattering lengths. The decay rate of the off-diagonal matrix element $|\rho_{a, b}^{\mathrm{el}}|$ is determined by both the real and imaginary parts of the scattering length. Eqs. (5) and (10) show that the effect of elastic scattering (the real parts of the scattering length) is contained in the coefficients proportional to $T^{1/2}$ or higher powers of temperature. In order to measure the effect of elastic scattering on collisional decoherence, it may therefore be necessary to raise the temperature of the gas. However, the collision system must remain in the $s$-wave scattering regime. The optimal temperature of the experiments must therefore be just below the temperature, at which the $p$-wave contirbution to the scattering cross sections becomes important. This temperature is different for different collision systems and depends on the reduced mass of the colliding particles, the strength of the inter-particle interaction, the strength of $s$-wave scattering and the magnitude of the resonant enhancement of the $s$-wave scattering cross sections.

\section{Conclusion}

In summary, we have demonstrated that collisional decoherence of ultracold molecules prepared in coherent superpositions of internal states can be controlled by tuning an external magnetic field near a scattering resonance. Our results show that the decoherence rate is suppressed when both the real and imaginary parts of the scattering lengths for the states in the coherent superposition are equal.  The scattering resonance modifies the scattering length of one of the states in the coherent superposition, which can be exploited to reduce or enhance the rate of collisional decoherence by varying the external field. 

The ability to tune collisional decoherence in a molecular gas by varying an external field can be used for numerous practical and fundamental applications. We have demonstrated that measuring the decay of the total and relative coherence signals can be used to determine the scattering length of molecules in any quantum state $x$, if the scattering length of one state is known. The proposed method relies on measuring the decay of coherence between a state that exhibits resonances and a state that is resonance-free.  Tuning collisional decoherence and measuring the coherence times can also be exploited for calibrating other less controllable sources of decoherence.  The role of quantum coherence in energy transport and diffusion in molecular gases remains an open question that can be explored in an experiment with tunable decoherence.  Quantum coherence may play an intriguing role in chemical dynamics of ultracold molecules. An experiment with tunable collisional decoherence might reveal novel features of quantum coherent chemistry.

\section{Acknowledgment}

This work is supported by NSERC of Canada. RVK acknowledges support from the Institute for Theoretical, Atomic, Molecular and Optical Physics at the Harvard-Smithsonian Center for Astrophysics in the form of a sabbatical fellowship. 

\begin{figure}
\begin{center}
\includegraphics[width=0.75\linewidth]{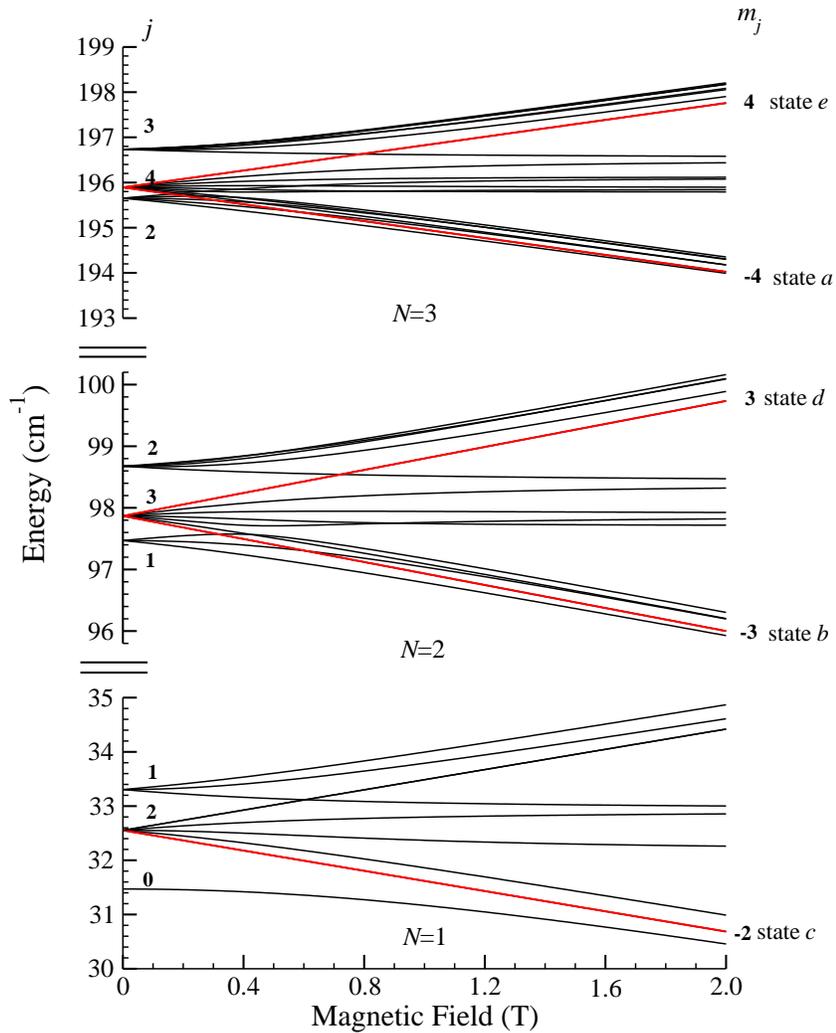}
\caption{(color online) Energy levels of the NH molecule in a magnetic field.}
\label{fig1}
\end{center}
\end{figure}

\begin{figure}
\begin{center}
\includegraphics[width=0.95\linewidth]{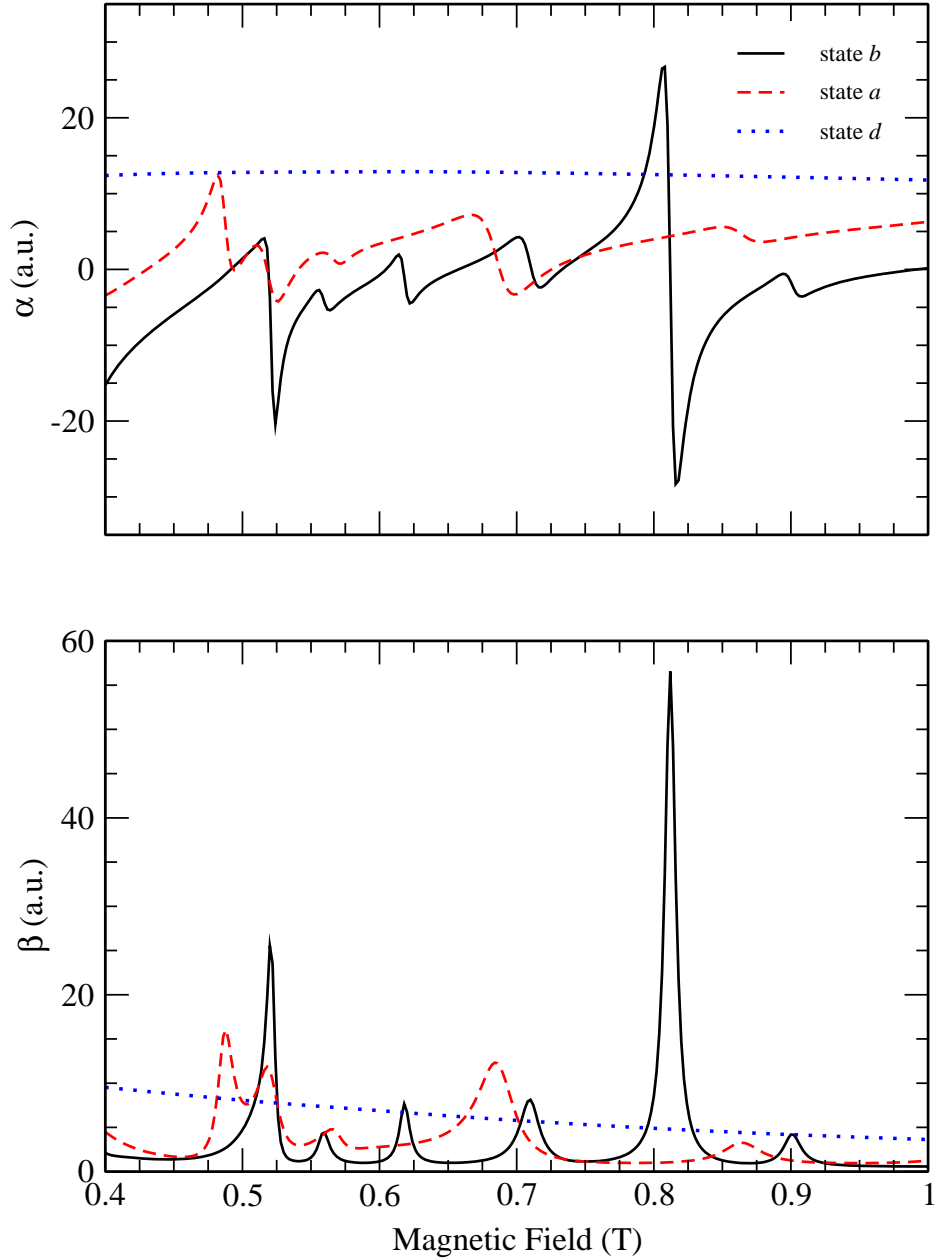}
\caption{(color online) The real (upper panel) and imaginary (lower panel) parts of 
the Mg -- NH $s$-wave scattering lengths: solid curves --  state $a$; dashed curves -- state $b$; dotted curves -- state $d$ of Figure 1. }
\label{fig1}
\end{center}
\end{figure}

\begin{figure}
\begin{center}
\includegraphics[width=0.95\linewidth]{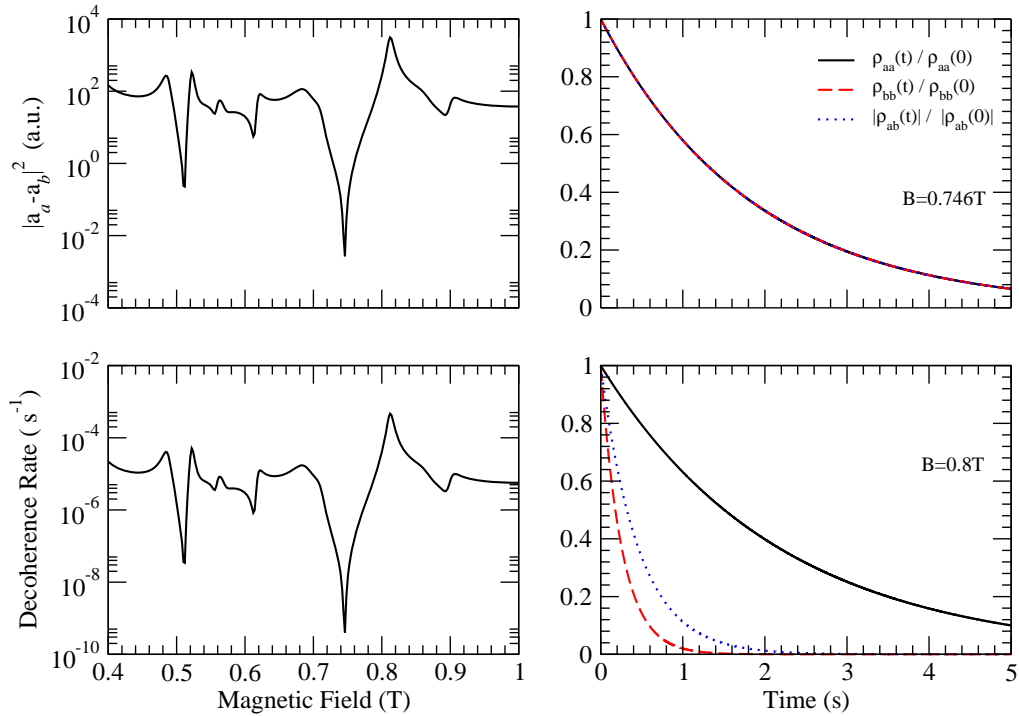}
\caption{(color online) Upper left panel: The absolute difference of the scattering lengths for NH molecules in states $a$ and $b$ (labeled in Figure 1) as a function of the magnetic field. 
Lower left panel: The $t=0$ decoherence rate of NH molecules prepared in the superposition of states $a$ and $b$  as a function of the magnetic field. Right panels: The time dependence of the density matrix elements $\rho_{a, a}^{\mathrm{el}}$ (dashed lines), $\rho_{b, b}^{\mathrm{el}}$ (solid lines) and $|\rho_{a, b}^{\mathrm{el}}|$ (dotted lines) at two magnitudes of the magnetic field.}
\label{fig2}
\end{center}
\end{figure}

\begin{figure}
\begin{center}
\includegraphics[width=0.95\linewidth]{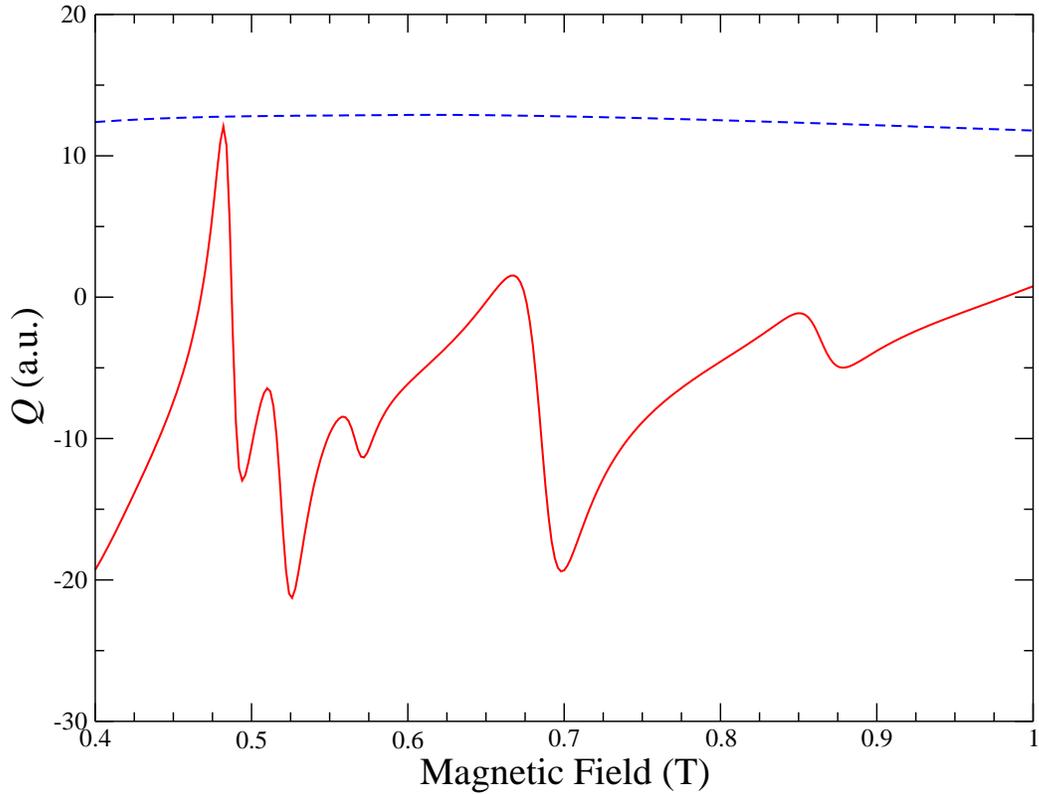}
\caption{The quantity $Q = \alpha_a \pm \sqrt{R_{ad}/CT^{1/2} - (\beta_a - \beta_d)^2}$: solid curve -- minus sign; dashed curve -- plus sign.
}
\label{fig2}
\end{center}
\end{figure}

%\begin{figure}
%\begin{center}
%\includegraphics[width=0.95\linewidth]{Figure4}
%\caption{The $t=0$ decoherence rate of NH molecules prepared in the superposition of states labeled in Figure 1 as a function of the magnetic field. The left panels -- states $b$ and $c$; the middle panels -- states  $a$ and $d$; the right panels -- states $b$ and $e$. The black lines in the upper panels depict the real part of the scattering lengths; the red lines -- the imaginary part of the scattering lengths. }
%\label{fig2}
%\end{center}
%\end{figure}

\end{document}